\documentclass[%
 aip,
 amsmath,amssymb,
 reprint,%
]{revtex4-1}

\usepackage{graphicx}
\usepackage{dcolumn}
\usepackage{bm}

\usepackage[utf8]{inputenc}
\usepackage[T1]{fontenc}
\usepackage{mathptmx}
\usepackage{etoolbox}
\usepackage{amssymb}
\usepackage{amsmath, nccmath}
\usepackage{mleftright}
\usepackage{siunitx}

\makeatletter
\def\@email#1#2{%
 \endgroup
 \patchcmd{\titleblock@produce}
  {\frontmatter@RRAPformat}
  {\frontmatter@RRAPformat{\produce@RRAP{*#1\href{mailto:#2}{#2}}}\frontmatter@RRAPformat}
  {}{}
}%
\makeatother
\begin{document}

\preprint{AIP/123-QED}

\title[]{Revised Fowler-Dubridge model for photoelectron emission from two-dimensional materials}

\author{Yi Luo}
\thanks{Authors to whom correspondence should be addressed: 
yi\_luo@sutd.edu.sg
ricky\_ang@sutd.edu.sg}
\affiliation{Science, Mathematics and Technology (SMT), Singapore University of Technology and Design (SUTD), 8 Somapah Road, Singapore 487372}

\author{Yee Sin Ang}
\affiliation{Science, Mathematics and Technology (SMT), Singapore University of Technology and Design (SUTD), 8 Somapah Road, Singapore 487372}

\author{L. K. Ang}
\thanks{Authors to whom correspondence should be addressed: 
yi\_luo@sutd.edu.sg
ricky\_ang@sutd.edu.sg}
\affiliation{Science, Mathematics and Technology (SMT), Singapore University of Technology and Design (SUTD), 8 Somapah Road, Singapore 487372}


\begin{abstract}
We revise the Fowler-Dubridge (FB) model for photoelectron emission from two-dimensional (2D) materials to include the effects of reduced dimensionality, non-parabolic and anisotropic energy dispersion of 2D materials.
Two different directions of electron emission are studied, namely vertical emission from the surface and lateral emission from the edge.
Our analytical model reveals a universal temperature scaling of $T^\beta$ with $\beta$ = 1 and $\beta$ = 3/2, respectively, for the surface and edge emission over a wide class of 2D materials, which are distinct from the traditional scaling of $\beta$ = 2 originally derived for the traditional bulk materials.
Our comparison shows good agreement to two experiments of photo-electron emitted from graphene for both surface and edge emission.
Our calculations also show the photoelectron emission is more pronounced than the coexisting thermionic emission for materials with low temperature and Fermi energy.
This model provides helpful guidance in choosing proper combinations of light intensity, temperature range and type of 2D materials for the design of photoemitters, photodetectors and other optoelectronics.
\end{abstract}

\maketitle

Photoelectron emission plays important roles in many research areas ranging from ultrafast microscopes, \cite{muller2014femtosecond,sun2020direct} particles accelerators, \cite{england2014dielectric} high current photocathodes \cite{polyakov2013plasmon,schroder2015ultrafast,li2019extreme} to nano-vacuum electronic devices. \cite{forati2016photoemission,lin2017electric,zhang2017100,zhou2021ultrafast,zhang2021space} 
Recently, there has been substantial interests in studying photo-excited electron emission from 2D materials, \cite{copuroglu2014gate,massicotte2016photo,higuchi2017light,son2018ultrafast,heide2019interaction,rezaeifar2019hot,madas2019superior,ahsan2020performance,heide2020attosecond} due to the unique properties of 2D materials, such as excellent mechanical flexibility, artificially stacking configuration, \cite{guo2021stacking} large carrier mobility, \cite{mir2020recent,du2008approaching} high temperature stability, \cite{van2019thermal,david2014evaluating} and good electrical and optical conductivity. \cite{zhou2019optical,weiss2012graphene}

These developments will be useful in the advancements of photoemitters \cite{yamaguchi2017active,son2018ultrafast,SJ2022a} and photodetectors. \cite{lopez2013ultrasensitive,long2019progress,qiu2021photodetectors,SJ2022b,zha2022infrared} 
For instance, Copuroglu \textit{et al}. \cite{copuroglu2014gate} explored the tunability of photoemission spectra from graphene via the change of gate bias. 
Higuchi \textit{et al}. \cite{higuchi2017light} showed the carrier-envelope-phase-controlled photoemission current in graphene under strong laser excitation, due to the breaking of inversion symmetry by incident few-cycle pulses.
Rezaeifar \textit{et al}. \cite{rezaeifar2019hot} presented efficient photoemission from a waveguide-integrated graphene using greatly reduced laser power and sub-work-function photons, which is attributed to the significant enhancement of hot electrons emission in the graphene.  
Madas \textit{et al}. \cite{madas2019superior} reported the superior photo-thermionic emission from illuminated anisotropic black phosphorene surface, which is found to be higher than that from graphene at same conditions. 

The simplest photoelectron emission model is known as the Fowler-Dubridge (FD) model. \cite{fowler1931analysis,dubridge1933theory,bechtel1975four} Unlike the field emission (or electron tunneling under the potential barrier) known as the Fowler-Nordheim (FN) model, \cite{fowler1928electron} FD model describes the over-barrier emission of electrons after absorption of photons, which is similar to the thermionic emission at high temperature.
Compared to the revised thermionic and field emission models developed for 2D materials, \cite{ang2017theoretical,ang2018universal,ang2019generalized,ang2021physics,chan2021thermal,chan2022field,ang2023universal}
the revision of photoelectron emission or FD model has not been studied.
Thus, in this paper, we develop analytical models for photo-excited over-barrier electron emission from either the edge or surface of 2D materials, where the reduction of dimensionality and non-parabolic and anisotropic energy dispersion of 2D materials are considered. 

Our analytical model reveals the unconventional scaling relation between the photoemission current and temperature for the edge and surface emission, which are different from that of traditional bulk materials. 
The power dependence of emission current on light intensity is recovered in the formulation, which provides a useful tool to explain the the experimentally measured photocurrent from monolayer graphene. \cite{heide2019interaction,rezaeifar2019hot}
Our model also  identifies the relative dominance of photoelectron emission over the coexisting thermionic emission process at different operating conditions.
This analytical model provides a theoretical avenue to design and analyze 2D-materials based opto-electronics involving photoelectron emission.


We first consider the photoelectron emission vertically from the surface of a 2D material 
that lies in the $x-y$ plane (cf. Fig. 1). 
Due to the reduced dimensionality along the vertical $z$ direction, the number of electrons for the over-barrier emission after absorbing photon energy per unit area inside the material is \cite{ang2018universal,ang2023universal}
\begin{equation}
        N_{as}=\frac{g}{(2\pi)^2}\int[f_{FD}(\varepsilon_{\parallel})\Gamma(\varepsilon_{\parallel})] d^2\textbf{k}_{\parallel},
\label{eq1}
\end{equation}
where $g$ is the spin-valley degeneracy factor, $\textbf{k}_{\parallel}=(k_x,k_y)$ and $\varepsilon_{\parallel}$ are the electron wavevector and energy component in the $x-y$ plane, respectively, $f_{FD}(\varepsilon_{\parallel})$ is the Fermi-Dirac distribution function, $\Gamma(\varepsilon_{\parallel})=H(\varepsilon_{\parallel}+n\hbar\omega-W_0)$ is the transmission probability, where $H(x)$ is the Heaviside function.
The other parameters: $\hbar$ is the reduced Plank constant, $\omega$ is the incident laser frequency, $W_0=\varepsilon_{F}+W$ with $\varepsilon_{F}$ being the Fermi energy and $W$ being the work function of material.
Here, electrons are assumed to be free and obey the Fermi-Dirac distribution, which is valid for the materials with conduction band easily populated, \cite{westover2010photo} such as graphene, 2D metallic film and some 2D semiconductors.
These electrons at energy $\varepsilon_{\parallel}$ will absorb $n$ photons to overcome the material-vacuum interface barrier $W_0$ to be emitted from the surface.

\begin{figure}[t]
\centering %
\includegraphics[width=0.48\textwidth]{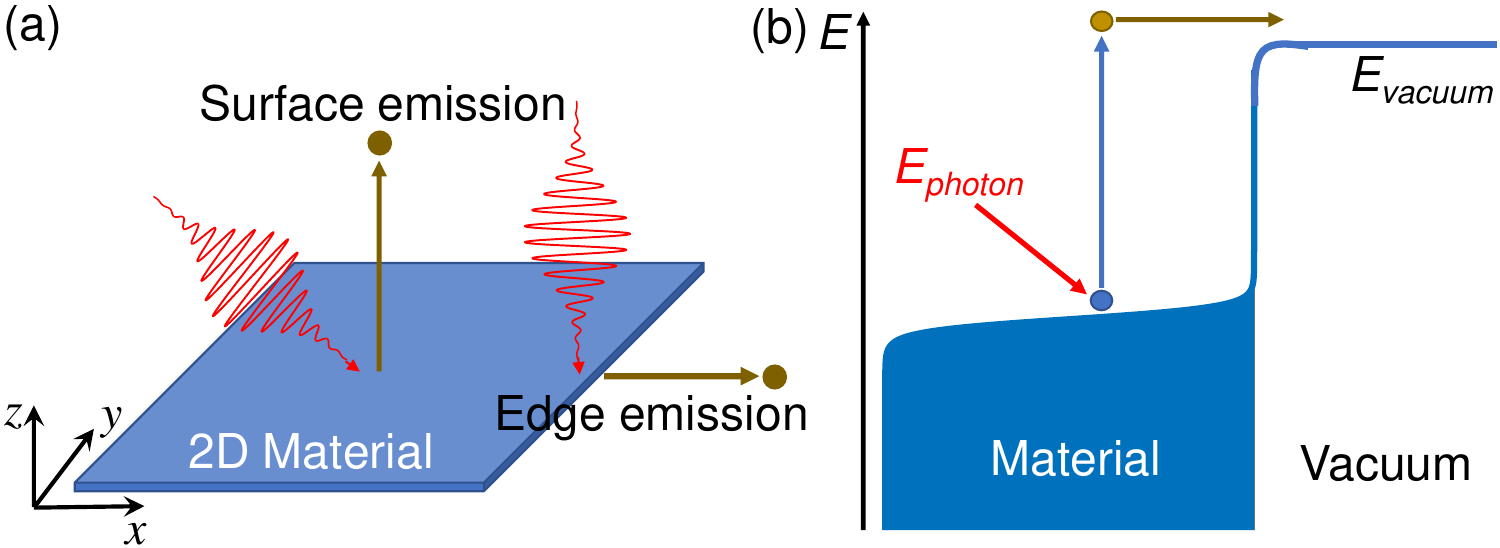} 
\caption{(a) Sketch of laser-induced electron emission from the surface and edge of two-dimensional (2D) materials. (b) Energy diagram for photon-driven over-barrier electron emission. Electron in the material is excited to emit via absorbing $n$ photon of energy $E_{photon}$.}
\label{fig1}
\end{figure}

To solve Eq. (\ref{eq1}), we implement the transformation of $gd^2\textbf{k}_{\parallel}/(2\pi)^2=D(\varepsilon_{\parallel})d\varepsilon_{\parallel}$, where $D(\varepsilon_{\parallel})$ is the electronic density of states (DOS), and a general form of 2D anisotropic DOS is assumed: \cite{ang2023universal} $D(\varepsilon_{\parallel})=g \times C_{k,q} \times \varepsilon_{\parallel}^{k}$ with $C_{k,q}$ being a material-dependent parameter and $|k|\leq 1$. 
For instance, the commonly used 2D materials for electron emission: monolayer graphene \cite{neto2009electronic} $D(\varepsilon_{\parallel})=g\varepsilon_{\parallel}/{2\pi\hbar^2 v_{F}^{2}}$, highly doped black phosphorous \cite{jiang2015magnetoelectronic} $D(\varepsilon_{\parallel})=g\sqrt{m_{cx}^{*} m_{cy}^{*}}/{2\pi\hbar^2}$, and $ABC$-stacked $N$-layer graphene \cite{ang2018universal,min2008electronic} $D(\varepsilon_{\parallel})=gt_{\perp}^{4/3}\varepsilon_{\parallel}^{-1/3}/{2\pi\hbar^2 v_{F}^{2}}$, all follow the general DOS expression above with different values of $k$. 
This assumed DOS allows our model to describe a large class of 2D materials and the calculated results can be formulated in analytical expression. 
Density functional theory (DFT) simulation may also be used to obtain the energy dispersion near the Fermi energy level in order to determine the values of $C_{k,q}$ and $k$ via appropriate numerical fitting.

Based on the DOS, Eq. (\ref{eq1}) becomes
\begin{equation}
        N_{as}=g \times C_{k,q} \int_{W_{0}-n\hbar\omega}^{+\infty} \frac{\varepsilon_{\parallel}^{k}}{\text{exp}[(\varepsilon_{\parallel}-E_{F})/k_{B}T]+1}d\varepsilon_{\parallel}.
\label{eq2}
\end{equation}
By making the substitution of $\mu=\varepsilon_{\parallel}/k_{B}T-\mu_{0}$ and $\mu_{0}=W_{0}/k_{B}T$, and performing a Taylor expansion of $\mu\ll\mu_{0}$ (see the Supplemental Materials for derivation details), Eq. (\ref{eq2}) can be analytically solved as 
\begin{fleqn}
\begin{equation}
\begin{aligned}
    N_{as}
    & =g \times C_{k,q}W_{0}^{k}k_{B}T\{ (1-k\frac{n\hbar\omega}{W_{0}})\text{In}[1+\text{exp}(\frac{n\hbar\omega-W}{k_{B}T})]- \\
    & \hspace{0.5cm} k\frac{k_{B}T}{W_{0}}Li_{2}[-\text{exp}(\frac{n\hbar\omega-W}{k_{B}T})]\},
\label{eq3}
\end{aligned}
\end{equation}
\end{fleqn}
where $Li_{2}(x)$ is the polylogarithm function of order 2. 
To verify the validity of Eq. (3), we compare it with the full numerical calculation in solving Eq. (\ref{eq2}) as a function of temperature $T$ for monolayer graphene, highly doped black phosphorous and $ABC$-stacked trilayer graphene in Fig. 2(a), which shows good agreements between them.
At low temperature regime of $W_{0}\gg k_{B}T $, Eq. (3) reduces to
$N_{as}= gC_{k,q}W_{0}^{k} \times k_{B} 
T \times (1-kn\hbar\omega/{W_{0}}) \times \text{In}[1+\text{exp}(n\hbar\omega-W)/(k_{B}T)]$, 
which displays a universal temperature $T^\beta$ with $\beta$ = 1 scaling in the prefactor and it is independent of type of 2D materials. 
This is in stark contrast to the $T^2$ scaling for traditional material exhibited in the original FD model. \cite{bechtel1975four} 

\begin{figure}[b]
\centering %
\includegraphics[width=0.48\textwidth]{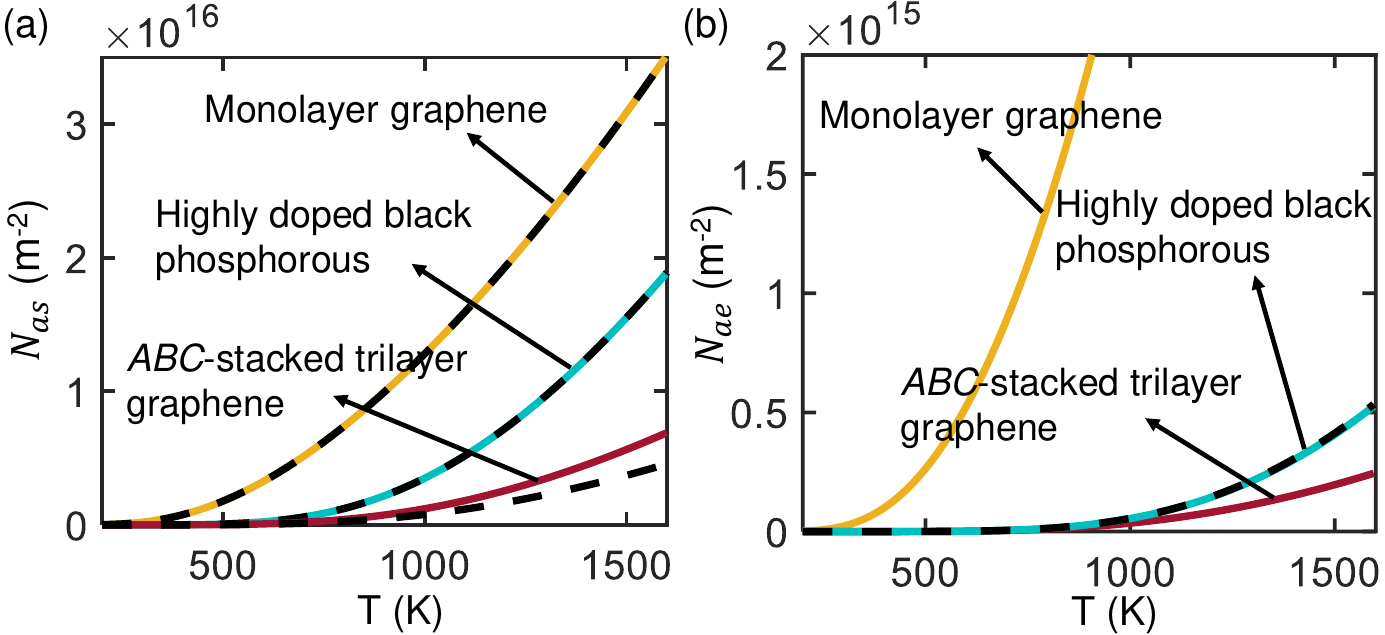}
\caption{Number of electrons per area inside 2D materials of photoelectron over-barrier emission for monolayer graphene, highly doped black phosphorous and $ABC$-stacked trilayer graphene: (a) surface emission $N_{as}$ and (b) edge emission $N_{ae}$. 
Solid lines denote the numerical solutions of Eq. (2) in (a) and Eq. (6) in (b). 
Dashed lines denote the analytical results from Eq. (3) in (a) and Eq. (7) in (b). 
For monolayer graphene, we have $g=4$, the Fermi level $\varepsilon_{F}$ = 0.2 eV, work function $W$ = 4.5 eV, and Fermi velocity $v_{F}\approx 10^{6}$ m/s.
For highly doped black phosphorous, $g=2$, $\varepsilon_{F}$ = 4.3 eV, $W$ = 4.56 eV, and effective masses are $m_{cx}^{*}\approx0.2m$ and $m_{cy}^{*}\approx0.4m$, where $m$ is the free-electron mass.
For $ABC$-stacked trilayer graphene, $g=4$, $\varepsilon_{F}$ = 0.2 eV, $W$ = 4.7 eV, $v_{F}\approx 10^{6}$ m/s and the interlayer hopping parameter $t_{\perp}\approx$ 0.39 eV. 
In all calculations, the incident laser wavelength is set at 800 nm with $n$ = 2.7 multiphoton absorption. 
}
\label{fig2}
\end{figure}

\begin{figure*}[t]
\centering
\includegraphics[width=0.9\textwidth]{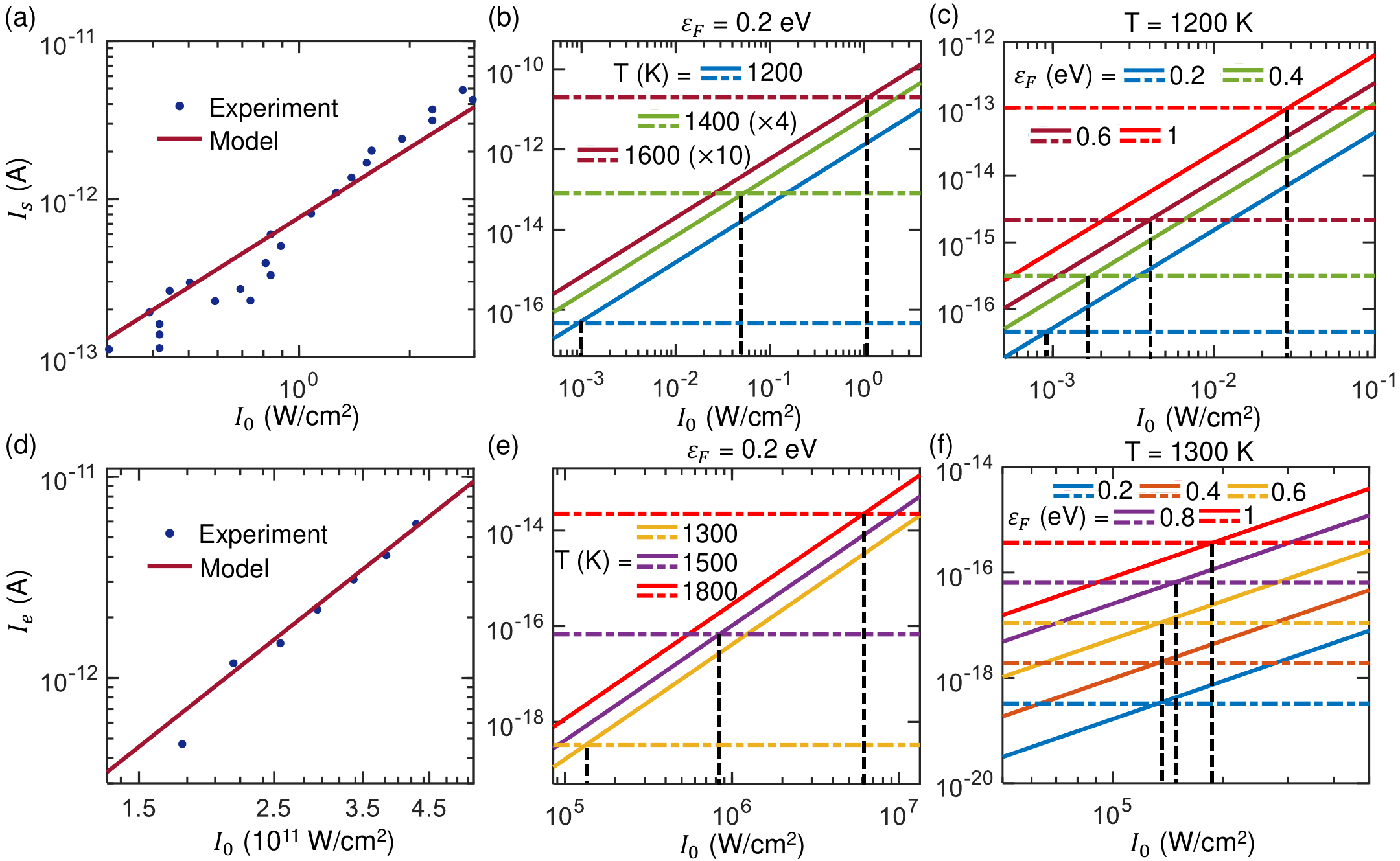}
\caption{Photoelectron emission current as a function of laser peak intensity $I_0$ from the surface [(a)-(c)] and edge [(d)-(f)] of a monolayer graphene. 
Comparison [(a),(d)] between our models (red solid lines) and two experiments (blue circles): surface emission (top) \cite{rezaeifar2019hot} and edge emission (bottom) \cite{heide2019interaction}. 
Comparison of calculations between the over-barrier photoemission current (solid lines) and thermionic emission current (dash-dotted lines) at different temperatures $T$ [(b),(e)] and different Fermi energies $\varepsilon_{F}$ [(c),(f)]. 
In (a)-(c), the laser wavelength is 405 nm, and emitting area is 402 $\SI{}{\um}^2$ from Ref. \cite{rezaeifar2019hot}.
In (d)-(f), the laser wavelength is 800 nm, and emitting size is 2 $\SI{}{\um}$ from Ref. \cite{heide2019interaction}.}
\label{fig3}
\end{figure*}

According to the FD model, \cite{fowler1931analysis,dubridge1933theory,bechtel1975four} the number of photoemitted electron per area per unit time is proportional to the product of probability of absorbing $n$ photons per area per time (i.e., $[I(1-R)/\hbar\omega]^n$) and the total number of electrons $N_{as}$, thus the photoemission current density is written as
\begin{equation}
        J_{sn}\propto e \times [\frac{I(1-R)}{\hbar\omega}]^n \times N_{as},
\label{eq4}
\end{equation}
where $e$ is the elementary charge, $R$ is the optical reflection coefficient, $I$ is the incident laser intensity and $N_{as}$ is given by Eq. (3). 
To include other effects not considered in Eq. (\ref{eq4}), such as electron band transition, scatterings and reflection at the interface, a material-dependent coefficient $a_n$ can be introduced, which needs to be determined by fitting to experimental measurement as suggested in the traditional FD model.

Inserting Eq. (3) into Eq. (\ref{eq4}), the vertical photoelectron emission current density from the surface of 2D materials is expressed in the FD-like form of 
\begin{fleqn}
\begin{equation}
\begin{aligned}
    J_{sn}
    & =a_{n} \times 
    gC_{k,q}W_{0}^{k}(\frac{e}{\hbar\omega})^{n}A(1-R)^{n}I_{0}^{n}k_{B}T\{ (1-k\frac{n\hbar\omega}{W_{0}})\times \\
    & \hspace{0.5cm} \text{In}[1+\text{exp}(\frac{n\hbar\omega-W}{k_{B}T})]-k\frac{k_{B}T}{W_{0}}Li_{2}[-\text{exp}(\frac{n\hbar\omega-W}{k_{B}T})]\},
\label{eq5}
\end{aligned}
\end{equation}
\end{fleqn}
where $A$ is the Richardson constant and $I_0$ denotes the laser intensity peak. 
Notably, Eq. (5) accounts for various $n$th-order photoemission processes. For instance, at $n = 0$ (i.e., pure thermionic emission), Eq. (5) reduces to the Richardson-Dushman thermionic emission model: $J_{s0}=a_{0}A \times k_{B}T \times \text{exp}(-W/k_{B}T)$ for $W\gg k_{B}T$, where the $T$ scaling is consistent with that of recently developed 2D thermionic surface emission model. \cite{ang2018universal} 
For general $n \neq 0$, Eq. (5) describes the thermally assisted photoelectron emission current density from the surface of the 2D materials as firstly reported in this paper.

We next consider the case of over-barrier emission from the edge of 2D material along $x$ direction (cf. Fig. 1). 
A general form of anisotropic energy dispersion in the 2D plane: $\varepsilon_{\parallel}(k_{x},k_{y})=(\alpha_{l}k_{x}^{2}+\beta_{l}k_{y}^2)^{l/2}$ is employed, where $\alpha_{l}$ and $\beta_{l}$ are the material-dependent parameters along $x$ and $y$ directions respectively, and $l$ is positive integer. 
The number of electrons available for the over-barrier edge emission per unit area becomes \cite{ang2018universal,ang2019generalized}
\begin{equation}
\begin{split}
        N_{ae}
        & =\frac{g}{(2\pi)^2}\int[f_{FD}(\varepsilon_{\parallel})\Gamma(\varepsilon_{x})] d^2\textbf{k}_{\parallel}\\
        &=\frac{g}{(2\pi)^2}\int_{\alpha_{l}^{l/2}k_{x}^{l}\geq W_{0}-n\hbar\omega}^{+\infty} dk_{x} \int_{-\infty}^{+\infty} \frac{dk_{y}}{\text{exp}[(\varepsilon_{\parallel}-E_{F})/k_{B}T]+1},
\label{eq6}
\end{split}
\end{equation}
where we have used the transmission probability $\Gamma(\varepsilon_{x})=H(\alpha_{l}^{l/2}k_{x}^{l}+n\hbar\omega-W_{0})$. 
Here, we assume that all the absorbed photon energy is converted into electron energy component along $x$ direction, and an electron can escape from the edge only if the $x$ component of electron energy is greater than the material-vacuum potential barrier. 
It is found that for $l=2$ type 2D materials, Eq. (\ref{eq6}) can be analytically solved (see the Supplemental Materials), which gives
\begin{equation}
N_{ae}^{(l=2)}=-\frac{g \times (k_{B}T)^{3/2}}{4\pi^2\sqrt{2\alpha_{l}\beta_{l}(W_{0}-n\hbar\omega)}} Li_{3/2}[-\text{exp}(\frac{n\hbar\omega-W}{k_{B}T})],
\label{eq7}
\end{equation}
where $Li_{3/2}(x)$ denotes the polylogarithm function of order 3/2. 
Equation (\ref{eq7}) displays another scaling of $T^\beta$ ($\beta$ = 3/2) in the prefactor, which is different from the traditional $\beta$ = 2 for bulk materials, and it has been observed in thermionic edge emission. \cite{ang2018universal} 
Note this $\beta$ = 3/2 (edge emission) is different from $\beta$ = 1 (surface emission) as described above. 

Similar to Eq. (5), the photoelectron emission current density from the edge of general 2D material is expressed as,
\begin{equation}
        J_{en}=a_n \times A(\frac{e}{\hbar\omega})^n(1-R)^nI^n \times N_{ae},
\label{eq8}
\end{equation}
where $N_{ae}$ is given by Eq. (\ref{eq6}). 
Note $N_{ae}$ of $l\neq2$ is not able to be well fitted with the analytical expression Eq. (7) (cf. Fig. 2(b)) or solved analytically.
In this work, we perform the numerical calculation of Eq. (\ref{eq6}) for some specific $l$: monolayer graphene ($l=1$) and $ABC$-stacked trilayer graphene ($l=3$).

Figure 3 shows the calculated photoemission current density as a function of laser peak intensity $I_0$ for monolayer graphene ($l$= 1) for both surface emission (top) and edge emission (bottom).
To obtain the value of empirical coefficients $a_n$ and $n$ in our models (cf. Eq. (5) and Eq. (8)), we fit (at room temperature $T$ = 300 K) the experimentally measured photocurrent of graphene for the surface emission \cite{rezaeifar2019hot} and edge emission \cite{heide2019interaction}, as shown in Figs. 3(a) and 3(d) respectively. 
Our model provides good fittings by using $a_n=3.6\times10^{-31}$ $\text{m}^2\text{K}^2(\text{m}^2\text{s}/\text{C})^{n}$ and $n =1.467$ in Fig. 3(a) and $a_n=3.83\times10^{-55}$ $\text{m}^3\text{K}^2(\text{m}^2\text{s}/\text{C})^{n}$ and $n =2.4$  in Fig. 3(d), where unknown $(1-R)^n$ value is absorbed into $a_{n}$. 
Notably, compared to Fig. 3(b), relatively low laser intensity $I_0$ in Fig. 3(a) can achieve the efficient photoemission, which is because the need for photoelectrons to be transported to the emitting surface is eliminated, significantly enhancing the photoemission probability. \cite{rezaeifar2019hot} 

\begin{figure}[t]
\centering %
\includegraphics[width=0.485\textwidth]{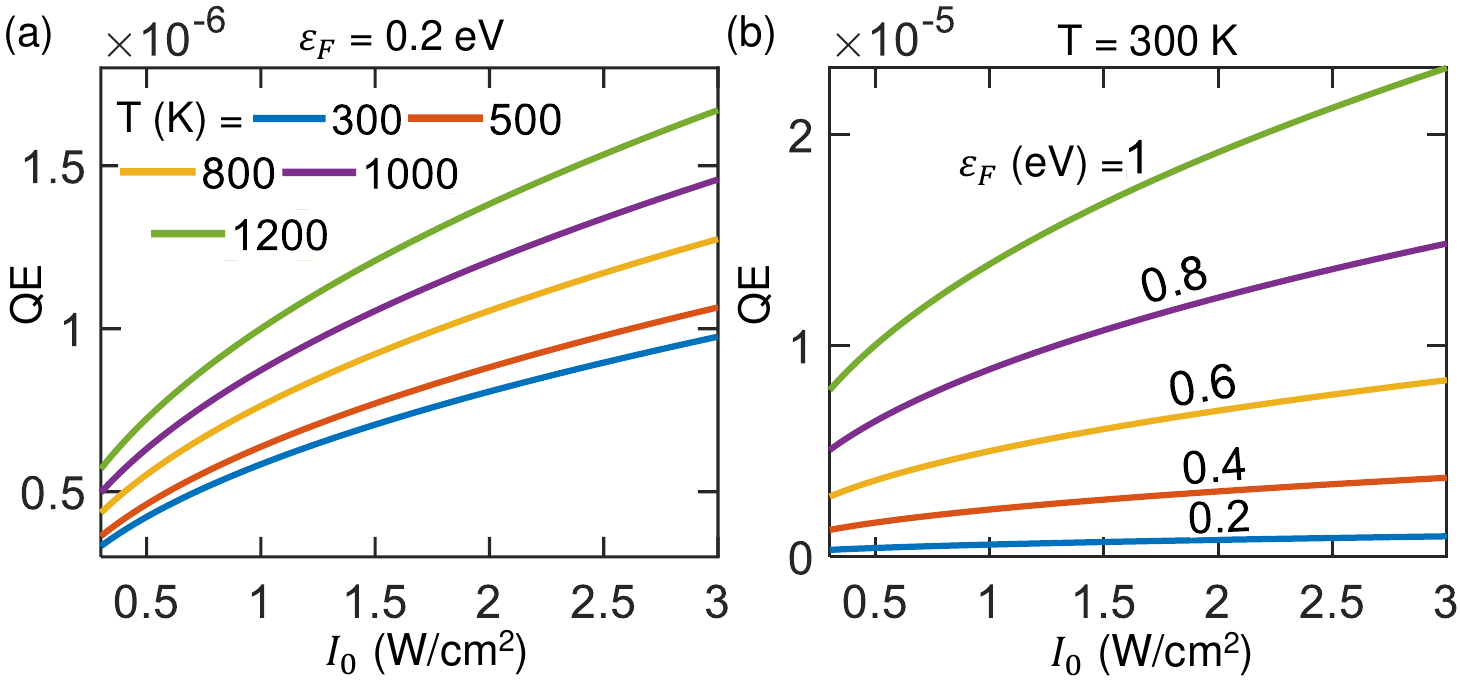} 
\caption{Quantum efficiency (QE) as a function of laser intensity $I_0$ from the surface of a monolayer graphene under different (a) temperatures $T$ and (b) Fermi energies $\varepsilon_{F}$. The value of empirical coefficients $a_n$ and $n$ in the model are extracted via fitting in Fig. 3(a). The incident laser wavelength is 405 nm. Emitting area is 402 $\SI{}{\um}^2$. \cite{rezaeifar2019hot}}
\label{fig4}
\end{figure}

We also plot the photoemission current (cf. solid lines) to compare with its coexisting pure thermionic emission current \cite{ang2018universal} (cf. dash-dotted lines) at different temperature ranges $T$ and Fermi energy levels $\epsilon_F$ = 0.2 to 1 eV in Figs. 3(b)-3(c) and 3(e)-3(f). 
Here, high temperature condition (at $T \geq 1200$ K) is found to have thermionic emission comparable to the photoelectron current in the laser intensity regime studied in Fig. 3. 
With the increasing temperature $T$, both over-barrier photoelectric emission and thermionic emission current increase, due to having more electrons available in the conduction band at higher $T$.
After a certain value of laser intensity (cf. black dashed lines), the over-barrier photoemission will become larger than the thermionic emission, manifesting the total current is mainly driven by the photo-excited electron emission. 
More importantly, this intensity threshold is greater at higher temperature $T$.
For example, for the surface emission (cf. Fig. 3(b)) at $T = 1200$ K, the photoemission-dominated region is above 0.001 W/cm$^2$ of laser intensity. At $T = 1400$ K, it moves to a higher value of 0.053 W/cm$^2$. 
This confirms photoelectron emission becomes more pronounced as compared to the thermionic emission at low temperature.

In Figs. 3(c) and 3(f), we study the effects of Fermi energy $\epsilon_F$ on the emission current as a function of laser intensity $I_0$. 
The tunability of Fermi level of graphene can be achieved through changing the doping concentration in graphene \cite{neto2009electronic} or external electric field strength. \cite{yu2009tuning}
It can be seen that over-barrier photoemission and thermionic emission current increase with the increasing Fermi energy $\varepsilon_{F}$, because more electrons are populated in the conduction band, providing higher availability of electron for the emission. 
Higher Fermi energy could shift the photocurrent-controlled emission region to the area at larger laser intensity threshold (cf. dashed lines in Figs. 3(c) and 3(f)), indicating the over-barrier photocurrent would be more noticeable than the thermionic emission at the lower Fermi energy.

Finally, Fig. 4 shows the quantum efficiency (QE = $J \times \hbar\omega/eI_0$) for photoelectron emission from the surface of monolayer graphene as a function of laser intensity $I_0$ at various temperatures $T$ in Fig. 4(a) and Fermi energies $\varepsilon_{F}$ in Fig. 4(b). 
The values of $a_n$ and $n$ are obtained from the experimental fitting in Fig. 3(a). 
Due to the order $n$ of multiphoton absorption being larger than 1, QE ($\propto I_{0}^{n-1}$) increases with intensity $I_0$ as shown in Fig. 4. 
Higher temperature or Fermi level can increase QE, because of larger photoemission current (cf. Figs. 3(b) and 3(c)). 
The increase of QE due to the higher temperature or Fermi energy is found to be more profound at stronger laser intensity.

In summary, we have presented a revised Fowler-Dubrigde (FD) model for photoelectron emission from either the surface or edge of 2D material, by explicitly considering the reduced dimensionality, non-parabolic and anisotropic energy band of a broad class of 2D materials.
Our analytical model reveals an universal temperature scaling of $T^{\beta}$ over a wide range of 2D materials: $\beta$ = 1 for surface emission, and $\beta$ = 3/2 for edge emission, which are different from the traditional $\beta$ = 2 scaling derived from the original FD model for 3D bulk materials. 
Based on a simple parametric analysis on monolayer graphene by using parameters from experimental measurements (surface emission\cite{rezaeifar2019hot} and edge emission \cite{heide2019interaction}), our model compares and identifies the dominance of photoelectron emission over the coexisting thermionic emission at different temperatures and Fermi energies.
This model may guide the design of photo-emitters or photo-detectors by choosing the suitable combinations of laser, temperature and 2D materials to be used in the applications.
Future work may include the various laser-matter interactions, such as laser induced non-equilibrium heating, \cite{WuPRB2008} finite pulse of the laser,\cite{PantPRB2012} and nonthermal photo-induced properties on quantum materials.  \cite{RMP.93.041002}\\

See the supplemental material for the derivation of analytical formulation of the number of electrons available for surface emission from 2D materials $N_{as}$ (Eq. (3)) and for edge emission from the parabolic 2D materials $N_{ae}^{(l=2)}$ (Eq. (7)).\\

This work is supported by the Singapore A*STAR IRG grant (A2083c0057). Y. S. A. is supported by the Singapore Ministry of Education (MOE) Academic Research Fund (AcRF) Tier 2 under Grant No. MOE-T2EP50221-0019.

\section*{AUTHOR DECLARATIONS}
\vspace{-0.5cm}
\subsection*{Conflict of Interest}
\vspace{-0.5cm}
The authors have no conflicts of interest to disclose.
\vspace{-0.7cm}
\subsection*{Author Contributions}
\vspace{-0.5cm}
\hspace{-0.35cm}\textbf{Yi Luo:} Formal analysis (lead); Investigation (lead); Writing-original draft (lead); Writing-review \& editing (equal). \textbf{Yee Sin Ang:} Conceptualization (equal); Investigation (equal); Supervision (equal); Wring-review \& editing (equal). \textbf{L. K. Ang:} Conceptualization (lead); Funding acquisition (lead); Investigation (equal); Supervision (Lead); Writing-review \& editing (equal).
\subsection*{DATA AVAILABILITY}
\vspace{-0.5cm}
The data that support the findings of this study are available from the corresponding author upon reasonable request.



\nocite{*}
\bibliography{biblio}

\end{document}


\maketitle

\section*{
\normalsize{\hspace{-0.17cm}I. Analytical formulation of the number of electrons available for surface over-barrier emission per unit area $N_{as}$}: Eq.(3) in the main text}
\vspace{-0.1in}
\hspace{0.1in} Here, we derive the analytical formulation of the number of electrons available for the surface over-barrier emission per unit area from 2D materials. Consider a general 2D anisotropic density of states (DOS) $D(\varepsilon_{\parallel})=gC_{k,q}\varepsilon_{\parallel}^{k}$ \cite{ang2023universal} with $C_{k,q}$ being material-dependent parameter and $|k|\leq 1$. Eq. (1) in the main text becomes 
\begin{equation}\tag{S1}
      N_{as}=gC_{k,q} \int_{W_{0}-n\hbar\omega}^{+\infty} \frac{\varepsilon_{\parallel}^{k}}{\text{exp}[(\varepsilon_{\parallel}-E_{F})/k_{B}T]+1}d\varepsilon_{\parallel},
\label{eqS1}
\end{equation}
where we have used the transformation $gd^2\textbf{k}_{\parallel}/(2\pi)^2=D(\varepsilon_{\parallel})d\varepsilon_{\parallel}$. 
By making the substitution of $\mu=\varepsilon_{\parallel}/k_{B}T-\mu_{0}$ and $\mu_{0}=W_{0}/k_{B}T$, Eq. (\ref{eqS1}) becomes
\begin{equation}\tag{S2}
\begin{split}
      N_{as}&=gC_{k,q} \int_{-n\hbar\omega/k_{B}T}^{+\infty} \frac{(k_{B}T)^{k+1} (\mu+\mu_{0})^k}{\text{exp}(\mu+\mu_{0}-E_{F}/k_{B}T)+1}d\mu \\
      & =gC_{k,q}(k_{B}T)^{k+1}\mu_{0}^{k} \int_{-n\hbar\omega/k_{B}T}^{+\infty} \frac{(1+\mu/\mu_{0})^k}{\text{exp}(\mu+\mu_{0}-E_{F}/k_{B}T)+1}d\mu.
\label{eqS2}
\end{split}
\end{equation}
By performing the Taylor expansion of $(1+\mu/\mu_{0})^k=1+k\mu/\mu_{0}+O(\mu^2/\mu^2_{0})$ for $\mu\ll\mu_{0}$ (valid in the nondegenerate emission regime \cite{ang2018universal}), Eq. (\ref{eqS2}) yields
\begin{equation}\tag{S3}
\begin{split}
      N_{as}& \approx gC_{k,q}(k_{B}T)^{k+1}\mu_{0}^{k} \int_{-n\hbar\omega/k_{B}T}^{+\infty} \frac{1+k\mu/\mu_{0}}{\text{exp}(\mu+\mu_{0}-E_{F}/k_{B}T)+1}d\mu \\
      &=gC_{k,q}(k_{B}T)^{k+1}\mu_{0}^{k}\{\text{In}[1+\text{exp}(\frac{E_{F}+n\hbar\omega}{k_{B}T}-\mu_{0})]-\frac{k}{\mu_{0}}\frac{n\hbar\omega}{k_{B}T}\text{In}[1+\text{exp}(\frac{E_{F}+n\hbar\omega}{k_{B}T}-\mu_{0})]- \\ 
      & \hspace{0.5cm} \frac{k}{\mu_{0}}Li_{2}[-\text{exp}(\frac{E_{F}+n\hbar\omega}{k_{B}T}-\mu_{0})]\} \\
      &=gC_{k,q}W_{0}^{k}k_{B}T\{ (1-k\frac{n\hbar\omega}{W_{0}})\text{In}[1+\text{exp}(\frac{n\hbar\omega-W}{k_{B}T})]-k\frac{k_{B}T}{W_{0}}Li_{2}[-\text{exp}(\frac{n\hbar\omega-W}{k_{B}T})]\}
\label{eqS3}
\end{split}
\end{equation}
which is Eq. (3) in the main text. \\ 

\section*{
\normalsize{\hspace{-0.17cm}II. Analytical formulation of the number of electrons available for edge over-barrier emission per unit area from the 2D material of $l$ = 2 $N_{ae}^{(l=2)}$}: Eq.(7) in the main text}
\vspace{-0.1in}
\hspace{0.1in} For $l=2$, Eq. (6) in the main text becomes
\begin{equation}\tag{S4}
\begin{split}
      N_{ae}^{(l=2)}
        &=\frac{g}{(2\pi)^2}\int_{\alpha_{l}k_{x}^{2}\geq W_{0}-n\hbar\omega}^{+\infty} dk_{x} \int_{-\infty}^{+\infty} \frac{dk_{y}}{\text{exp}[(\alpha_{l}k_{x}^{2}+\beta_{l}k_{y}^2-E_{F})/k_{B}T]+1} \\
        &=-\sqrt{\frac{\pi k_{B}T}{\beta_{l}}}\frac{g}{4\pi^2}\int_{\alpha_{l}k_{x}^{2}\geq W_{0}-n\hbar\omega}^{+\infty} Li_{1/2}[-\text{exp}(\frac{E_{F}-\alpha_{l}k_{x}^2}{k_B T})]dk_{x},
\label{eqS4}
\end{split}
\end{equation}
where $Li_{1/2}(x)$ denotes the polylogarithm function of order 1/2. Making the substitution $U=\alpha_{l}k_{x}^2-W_{0}+n\hbar\omega$ gives 
\begin{equation}\tag{S5}
      N_{ae}^{(l=2)}=-\frac{g}{4\pi^2\sqrt{2\alpha_{l}\beta_{l}}}k_{B}T\int_{0}^{+\infty} Li_{1/2}[-\text{exp}(\frac{E_{F}-W_{0}+n\hbar\omega}{k_B T}-U)]\frac{1}{\sqrt{U+(W_{0}-n\hbar\omega)/k_BT}}dU.
\label{eqS5}
\end{equation}
Here, since $1/\sqrt{U+(W_{0}-n\hbar\omega)/k_{B}T}$ tends to becomes zero as $U$ increases, the integrand of Eq. (S5) is essentially finite. We can assume the difference between total barrier height $W_{0}$ and the absorbed photon energy $n\hbar\omega$ (i.e., $W_{0}-n\hbar\omega$) is always large, leading to $(W_{0}-n\hbar\omega)/k_{B}T \gg U$. By neglecting $U$ in the denominator of integrand,  Eq. (S5) becomes 
\begin{equation}\tag{S6}
\begin{split}
      N_{ae}^{(l=2)}
        &=-\frac{g}{4\pi^2\sqrt{2\alpha_{l}\beta_{l}}}\frac{1}{\sqrt{W_{0}-n\hbar\omega}}(k_{B}T)^{3/2}\int_{0}^{+\infty} Li_{1/2}[-\text{exp}(\frac{n\hbar\omega-W}{k_B T}-U)]dU \\
        &=-\frac{g}{4\pi^2\sqrt{2\alpha_{l}\beta_{l}}}\frac{1}{\sqrt{W_{0}-n\hbar\omega}}(k_{B}T)^{3/2} Li_{3/2}[-\text{exp}(\frac{n\hbar\omega-W}{k_B T})],
\label{eqS6}
\end{split}
\end{equation}
which is Eq. (7) in the main text.

\bibliographystyle{aip}

\bibliography{biblio}